\documentclass{epl}
\usepackage{epsfig,psfrag}

\title{Resonant tunneling in a Luttinger liquid for arbitrary barrier 
transmission}
\shorttitle{Resonant tunneling in a Luttinger liquid}
\author{S. H\"ugle \and R. Egger}
\institute{Institut f\"ur Theoretische Physik, 
Heinrich-Heine-Universit\"at,
 D-40225 D\"usseldorf}
\pacs{71.10.Pm}{Fermions in reduced dimensions}
\pacs{73.23.Hk}{Coulomb blockade, single electron tunneling}
\pacs{73.40.Gk}{Tunneling}

\begin{document}

\maketitle

\begin{abstract}
A numerically exact dynamical quantum Monte Carlo approach has been
developed and applied to transport through a double barrier 
in a Luttinger liquid with arbitrary transmission. 
For strong transmission, we find broad Fabry-Perot Coulomb blockade
peaks, with a  lineshape parametrized by a single parameter, but at 
sufficiently low temperatures, non-Lorentzian universal lineshapes
characteristic of coherent resonant tunneling emerge, even for strong
interactions.  For weak transmission, our data supports the recently
proposed correlated sequential tunneling picture and 
is consistent with experimental results on intrinsic nanotube dots.
\end{abstract}

Resonant tunneling in a (non-chiral) Luttinger liquid (LL)
was studied more than a decade ago \cite{kf,furusaki,moon}, but
has recently attracted widespread 
attention by theorists again
\cite{furu98,kleimann1,braggio,thorwart,nazarov,polyakov,komnik}. 
This interest is primarily caused by new experimental 
realizations of double-barrier setups in interacting 1D
quantum wires presumably described by LL theory, using
semiconductor quantum wires \cite{auslaender} or nanotubes 
\cite{postma}.  The experiments of Ref.~\cite{postma} 
have been interpreted in terms
of a ``correlated sequential tunneling'' (CST)  mechanism
\cite{postma,thorwart}, since the standard picture of  
uncorrelated sequential tunneling (UST) in a LL \cite{furu98}
is inconsistent with the observed temperature dependence of the 
conductance peak height.   However, as CST theory essentially
relies on a master equation \cite{future2}, it is clearly of interest to 
check it against exact results. 
In this context also other nanotube experiments are of interest, where
for nearly transparent double-barrier,
Fabry-Perot oscillations in the gate-voltage dependence
of the conductance have been reported \cite{liang,kong,bockrath,park}. 

As the double-impurity problem in a LL is not integrable, 
exact solutions covering a wide parameter range
are out of reach, and  analytical progress 
has to rely on approximations.  One line of reasoning 
considered very weak Coulomb interactions, i.e.~Luttinger parameter
$g$ very close to one \cite{nazarov,polyakov}, where no 
CST processes were found.  
Furusaki has studied the UST regime, where the linewidth of the 
resonance peak has a linear temperature dependence, 
and the peak conductance $G_p\propto T^{-2+1/g}$ \cite{furu98}.
The CST mechanism could provide an additional transport channel
that may dominate on resonance for strong interactions, and gives
instead the observed \cite{postma}
behavior, $G_p\propto T^{-3+2/g}$ \cite{thorwart}. 
The CST mechanism is similar but different from conventional
cotunneling, which can be the dominant transport channel away 
from the resonance.  
We mention in passing that the exact $g=1/2$ solution of a modified 
model in Ref.\cite{komnik} neither contains the UST nor the CST 
regime and therefore cannot resolve this debate.
Finally, at low temperatures, instead of sequential tunneling, 
coherent resonant tunneling is possible, characterized by
non-Lorentzian universal lineshapes \cite{kf},
\begin{equation}\label{universal}
G(N_0,T)/G_0=f_g(X), \quad X=c |N_0-1/2|/T^{1-g}, 
\end{equation}
with dimensionful constant $c$ and $G_0=e^2/h$. Here
$N_0$ is the dot's average charge induced 
by external gates, with peak center, say, at $N_0=1/2$.  Obviously,
the linewidth then scales as $T^{1-g}$.
The universal scaling function has the limiting behavior
\cite{kf}
\begin{equation}\label{scallimit}
1-f_g(X\ll 1)\propto X^2,\quad 
f_g(X\gg 1)\propto X^{-2/g}.
\end{equation}
 While for
strong barriers, coherent resonant tunneling is only expected
for $g>1/2$ \cite{furu98}, for weak barriers (strong transmission), 
it is largely unclear to what extent this concept applies. 

Here we present numerically exact results obtained from
a dynamical quantum Monte Carlo approach that provide
detailed insight into the strong transmission regime 
and also sheds light on the controversy about mechanisms of
 resonant tunneling.
This method was successfully used for the corresponding
single-barrier case \cite{mak,leung}, and is generalized here to 
resonant tunneling.  We focus on the linear conductance 
for spinless electrons and symmetric barriers, since additional
simulations show that neither spin/flavor degeneracy 
nor weak asymmetry have a qualitative effect on our results.

We consider a LL containing two impurities of
strength $V_0$ at $x=\pm d/2$, thereby forming a quantum dot with LL leads.
The single-particle level spacing is $E_s=\pi v_F/d$,
the charging energy is  $E_c= E_s/ g^2$.  
In terms of the standard boson field $\phi(x)$ and its 
conjugate momentum $\Pi(x)$, the Hamiltonian is \cite{kf}
\begin{equation}
\label{hamg1}
H(t) =  \frac{v_F}{2} \int dx \ \left \{ \Pi^2 + \frac{1}{g^2} 
(\partial_x \phi)^2 \right \} 
+ V_0 \sum_{p=\pm}
\cos[\sqrt{4\pi} \phi(pd/2,t) + eVt +p\pi N_0 ],
\end{equation}
where $v_F$ is the Fermi velocity,
$V$ the applied bias voltage, and the 
current through the dot is 
\begin{equation} \label{cur}
I= G_0 V + \frac{e}{\sqrt{\pi}} \langle\partial_t \phi(x,t)\rangle,
\end{equation}
where $x$ is arbitrary and $t\to \infty$.
For $g=1$, refermionization yields the exact conductance
$G=dI/dV$.  With bandwidth $D$, the dimensionless linewidth 
\begin{equation} \label{paa}
w = \frac{(4-\lambda^2)^2}{8\lambda(4+\lambda^2)}, \quad \lambda = \pi V_0/D, 
\end{equation}
and the derivative of the Fermi function, $-df/dE= 1/[4T \cosh^2(E/2T)]$,
we obtain
\begin{equation}\label{cond1}
\frac{G(N_0,T)}{G_0} = \int_{-\infty}^\infty
 dE \frac{-df}{dE} \frac{ w^2}{\cos^2(\pi[N_0+E/E_s]) + w^2 },
\end{equation}
where $\hbar=k_B=1$.
For strong barriers, this leads to the usual Breit-Wigner lineshape
with linewidth $w E_s/\pi$. Note that
the infinite-barrier limit is reached already for $\lambda=2$, where
the associated phase shift is in the unitary limit.  
Equation (\ref{cond1}) holds for arbitrary barrier height $V_0$,
including strong transmission ($V_0\to 0$),
and allows to firmly establish the validity of our numerical scheme.

Next we  outline our path-integral Monte Carlo
(PIMC)  approach to the linear conductance for arbitrary $g$.  
While PIMC is conventionally used to evaluate imaginary-time path integrals, 
conductance calculations need dynamical information.
We first tried to use various schemes to analytically 
continue imaginary-time PIMC data, but the results were not
reliable.  This reflects a well-known difficulty related to the
numerically ill-posed nature of the analytic 
continuation \cite{leung}.  We therefore proceed directly within
a (Keldysh) real-time formalism.  Although real-time PIMC
has to deal with the sign problem, our formulation
avoids much of it by mapping the problem
to an equivalent Coulomb gas description.
In this representation, the sign problem is rather weak and permits
numerically exact simulations for
the full parameter regime of interest.

Consider the discretized Keldysh contour running from time $t=0$ 
to $t_{\rm max}$ and back to zero.
We keep  $t_{\rm max}$ finite and define a discrete time spacing 
$\Delta= t_{\rm max}/P$ with Trotter number $P$. 
At time $t_j=(j-1)\Delta$,
fields $\phi_{j}(x)$ ($\phi'_j(x)$) live on the forward (backward)
 branch.
Next we switch to a Coulomb gas picture by 
expanding the impurity propagator 
for sufficiently small $\Delta V_0$ \cite{leung}. Following Ref.~\cite{mak},
we use a Coulomb gas expansion valid up to order $(\Delta V_0)^2$.
Introducing  ``quantum'' charges $\xi_{jp}=0,\pm 1,\pm 2$ 
and ``quasiclassical''
charges $\eta_{jp}=0,\pm 1/2,\pm 1$ for each time 
 ($j=1,\ldots,P$) and impurity index ($p=\pm$), 
where $\eta\pm \xi/2$ must be integer,
it is sufficient to keep $|\eta+ \xi/2 | + 
| \eta-\xi/2 | \leq 2$ within this order of accuracy.
Only configurations subject to electroneutrality,
$\sum_{jp} \xi_{jp} = 0$, 
contribute to the partition function.
Moreover, it turns out that the quasiclassical $\eta$ 
charges can be summed over analytically \cite{mak,leung}.
 With $z$ defined in Eq.~(\ref{zdef}) below, 
this leads to effective Greens functions $K(\xi,z)$, with the entries
$K(0,z)=1-2(\Delta V_0/2)^2 (1-\cos z)$, 
$K(\pm 1,z) = \pm \Delta V_0 \sin(z/2)$, and
 $K( \pm 2,z) = (\Delta V_0/2)^2 (1-\cos z)$.
Under the Coulomb gas expansion, we can now integrate out all
boson fields $\phi_i(x)$ and $\phi'_i(x)$, since they appear
only quadratically in the action. 
Thereby we arrive at an effective action governing the 
dynamics of the Coulomb gas charges $\{\xi\}$.  The result can be
put into the language of dissipative quantum mechanics \cite{uli} 
by defining spectral densities
$J_{\pm}(\omega) = \pi g \omega [1\pm\cos (\pi \omega/E_c)] 
e^{-\omega/D}$,
with associated correlation function
\[
L_\pm(t) = 
 \int_0^\infty \frac{d\omega}{\pi} \frac{J_\pm(\omega)}{\omega^2}
 \frac{\cosh[\omega(1/2T-it)]-\cosh[\omega/2T]}{\sinh[\omega/2T]}.
\]
This implies (i) the action contribution 
$ \Phi' [\{\xi\}]= \sum_{pp'} \sum_{j\geq k}  
\xi_{jp} [S_{+,jk}+pp'S_{-,jk}] \xi_{kp'}, $
and (ii) the $z$'s entering $K(\xi,z)$ are  given by
\begin{equation}\label{zdef}
z_{kp} [\{\xi\}]= - 2 \sum_{j\geq k} 
\sum_{p'} \xi_{jp'} (R_{+,jk}+pp'R_{-,jk}),
\end{equation}
with $S_{\pm, jk}+ iR_{\pm, jk}=
L_\pm([j-k]\Delta)$; for the diagonal elements, see Ref.~\cite{JPC}.

Collecting results, the conductance is obtained as
\begin{equation}\label{cond}
G(N_0,T)/G_0 = 1 -\lim_{t\to \infty} \partial_t I_B(t),
\end{equation}
 where the function $I_B(t)$ can be computed from
\begin{equation}\label{IBK}
I_B(t_k)= Z^{-1} \sum_{\{\xi \}} 
I_k[ \{\xi\}] \cos\left(\pi N_0 \sum_{jp}p\xi_{jp}\right)
\exp(-\Phi'[\{\xi\}])  \prod_{jp} K(\xi_{jp},z_{jp}) .
\end{equation}
The normalization $Z$ is the path sum for $I_k\to 1$, and
\[
I_k [\{\xi\}]= \Delta \sum_{j'p'} j'\xi_{j'p'} \Bigl \{
\sum_{jp} S_{+,jk} \xi_{jp} 
+ 2\sum_{j\leq k,p} 
R_{+,kj} \frac{\partial_z K(\xi_{jp},z_{jp}) }{
K(\xi_{jp},z_{jp})}  \Bigr \}.
\]
This is a real-valued quantity, as are all the other quantities 
appearing in Eq.~(\ref{IBK}).
Remarkably, although we are dealing with a real-time sign problem, 
it effectively looks just like a fermion one.  The sign problem
arises because the integrand in Eq.~(\ref{IBK}) can be negative,
leading to interference effects and small signal-to-noise ratio
at sufficiently long real times $t_{\rm max}$.

\begin{figure}
\centerline{\epsfxsize=7.5cm \epsfysize=7cm \epsffile{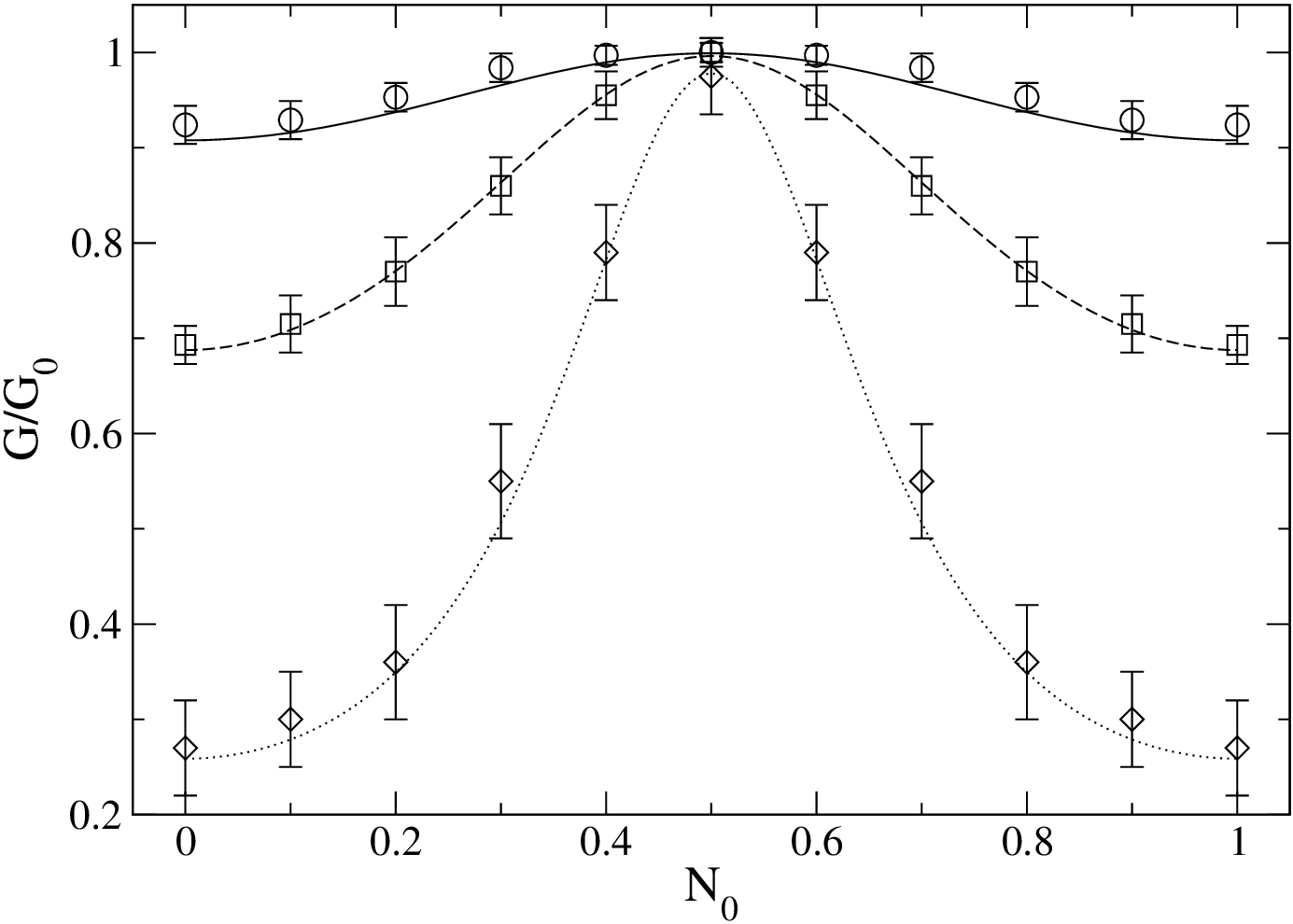}}
\caption{\label{fig1} Linear conductance versus $N_0$ for
 $g=1, E_s/D=\pi/2, T/D=0.025$,  with
$V_0/D=0.05$ (PIMC: circles; Eq.~(\ref{cond1}): solid curve), 
 $V_0/D=0.1$ (squares; dashed curve), 
 and $V_0/D=0.2$ (diamonds; dotted curve). 
}
\end{figure}

Monte Carlo trajectories for the quantum fluctuations $\{\xi\}$ 
are then generated by taking
the absolute value of the integrand in Eq.~(\ref{IBK}) as weight.
The average sign in the data reported here was always larger
than $\approx 0.001$, which still allows to run stable simulations,
albeit sometimes at the expense of very long CPU times. 
One can obtain the whole function $I_B(t)$ in one MC run.
For sufficiently long times, this function has a well-defined linear slope
which determines the conductance via Eq.~(\ref{cond}).  
Typically, Trotter convergence was reached for 
discretizations $\Delta V_0 \leq 0.1$. 
 On a 2 GHz Xeon processor our code
performs at an average speed of about $10^5$ samples per 
hour (for $P=40$). 
Several $10^6$ MC samples have to be
accumulated to obtain $I_B(t)$ for a given parameter set in order
to ensure good statistics. 
 Error bars then refer to
both standard stochastic MC errors 
and to uncertainties from fitting the long-time behavior by a linear slope.
The validity and accuracy of our scheme has been established by
 checking numerical data
against the exact $g=1$ solution, Eq.~(\ref{cond1}),
see Fig.~\ref{fig1}.   The comparison highlights
the power of our approach.
We then move on to interacting electrons, focusing
on $g=0.3$ and $g=0.6$.

\begin{figure}
\centerline{\epsfxsize=6.5cm \epsfysize=6cm \epsffile{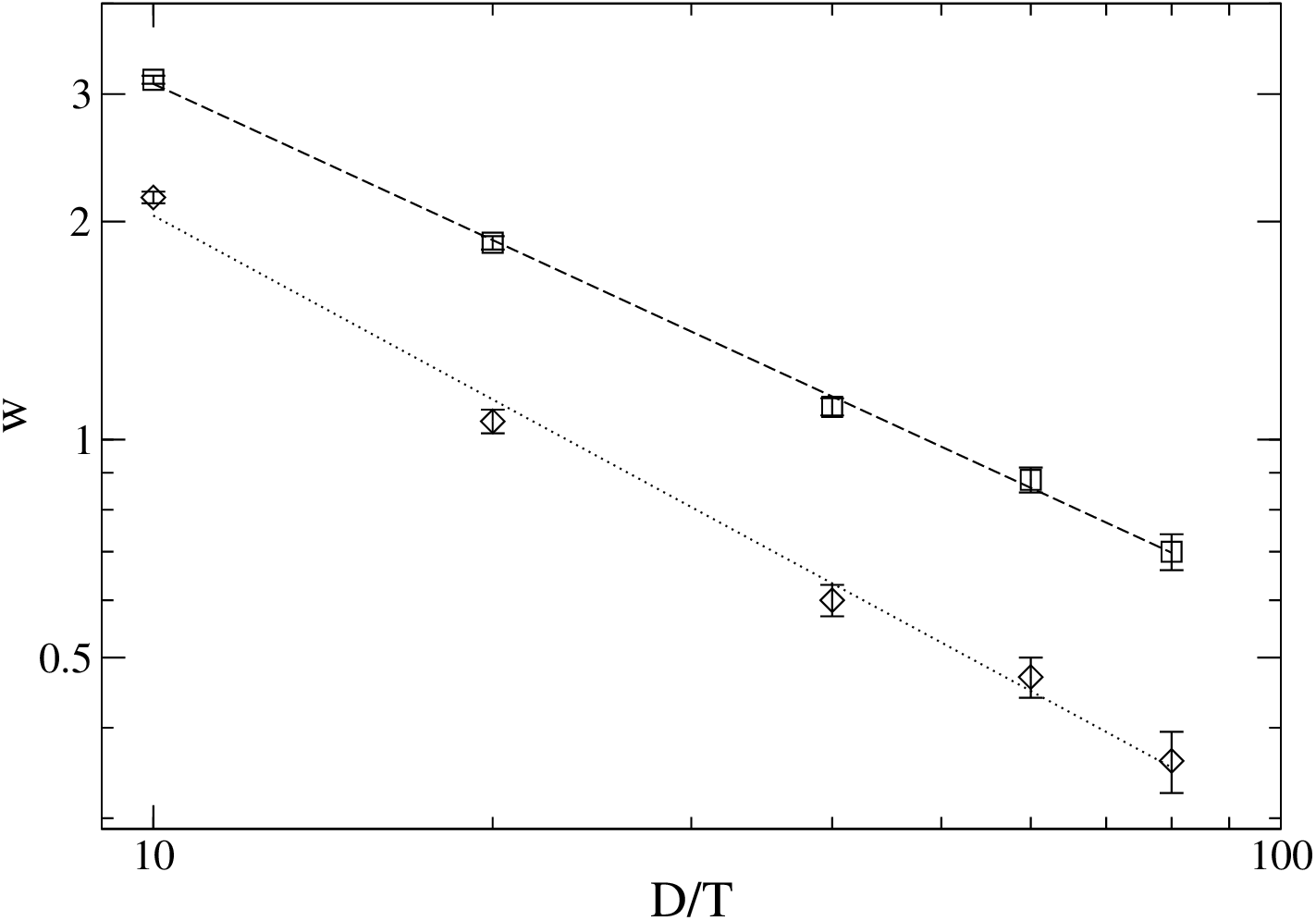}}
\caption{\label{fig2} Fit parameter $w_g(T)$ for conductance
lineshape (\ref{cond1}) as obtained from QMC data
at $g=0.3$ (diamonds) and
$g=0.6$ (squares), for $E_s/D=\pi/2$ and $V_0/D=0.05$. 
Dotted and dashed lines are guides to the eye only.
}
\end{figure}

Let us start with the case of strong transmission, taking 
$V_0/D=0.05$.
Remarkably, for $T/D >0.01$, PIMC data are {\sl quantitatively}
described by the $g=1$ lineshape (\ref{cond1}) provided $w=w_g(T)$
is treated as a fit parameter.  The corresponding values of $w$
are shown in Fig.~\ref{fig2}, and reveal power-law behavior, 
$w_g(T)\propto T^{\alpha_g}$, with $g$-dependent exponent 
$\alpha_g$.  For very weak interactions, a related behavior has been
discussed before \cite{polyakov}. Here we find
 numerical evidence for a power-law 
temperature dependence of the linewidth for strong interactions.
 For the two values of $g$ studied, we obtain
$\alpha_{0.3}\approx 0.84$ and $\alpha_{0.6}\approx 0.72$.
Therefore the strong-transmission
peaks become narrower and narrower as $T$ is lowered.
Note that each data point shown in Fig.~\ref{fig2} actually
has been obtained from QMC data for the full conductance lineshape.
The lineshape (\ref{cond1}) closely resembles experimental results 
for strong-transmission (Fabry-Perot) oscillations
 in nanotubes \cite{liang,kong,bockrath,park}.
We therefore identify this region of weak barriers and not too low
temperature as the {\sl Fabry-Perot regime}. 
For $g<1$, such Fabry-Perot oscillations include 
remnants of Coulomb blockade effects, which are partly washed 
out due to pronounced quantum fluctuations present at strong transmission.
Nevertheless, these effects are 
responsible for the narrowing of the resonance peak as temperature
is lowered.  We mention in passing that no narrow conductance {\sl dips} were
observed such as the ones seen experimentally in Ref.~\cite{liang}.
Such dips are probably related to special impurity scattering processes
not contained in the model (\ref{hamg1}).

\begin{figure}
\centerline{\epsfxsize=7.5cm \epsfysize=7cm \epsffile{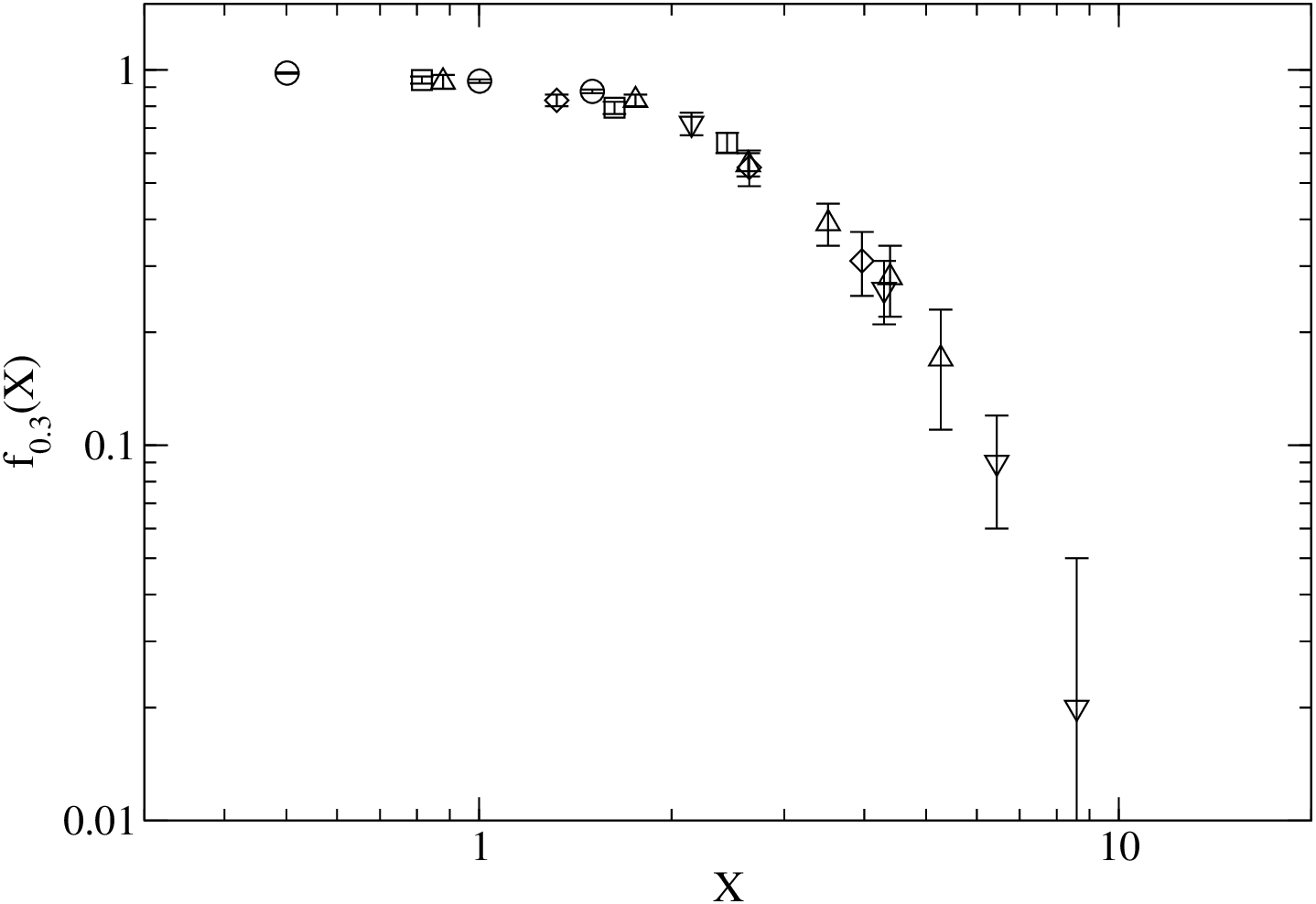}}
\caption{\label{fig3}
Low-temperature QMC data for various $N_0,T$
 at $E_s/D=\pi/2, V_0/D=0.05$, plotted according to Eq.~(\ref{universal}) 
for $g=0.3$.   Different symbols refer to different temperatures. 
}
\end{figure}

At lower temperatures, deviations from the Fabry-Perot lineshape
(\ref{cond1}) can be seen.  However, then our
 data can be collapsed onto the universal scaling
curve (\ref{universal}), see Fig.~\ref{fig3} for $g=0.3$.
Very similar results were found for $g=0.6$ as well.
For small and large $X=c|N_0-1/2|/T^{1-g}$, the results in 
Fig.~\ref{fig3} are consistent with the respective analytical
prediction (\ref{scallimit}). Therefore these universal lineshapes 
can be identified as {\sl coherent resonant tunneling} peaks.
Although for strong barriers, resonant tunneling exists only 
for $g>1/2$ \cite{kf}, we observe a perfect resonance peak at $g=0.3$.  This
is in accordance with renormalization group arguments for
weak impurities for $g>1/4$ \cite{kf}, and shows that the picture of
 coherent resonant tunneling in a Luttinger liquid is
actually very robust.  Only for very weak barriers and sufficiently high $T$,
the Fabry-Perot regime replaces the universal region of 
resonant tunneling.

\begin{figure}
\centerline{\epsfxsize=7.5cm \epsfysize=7cm \epsffile{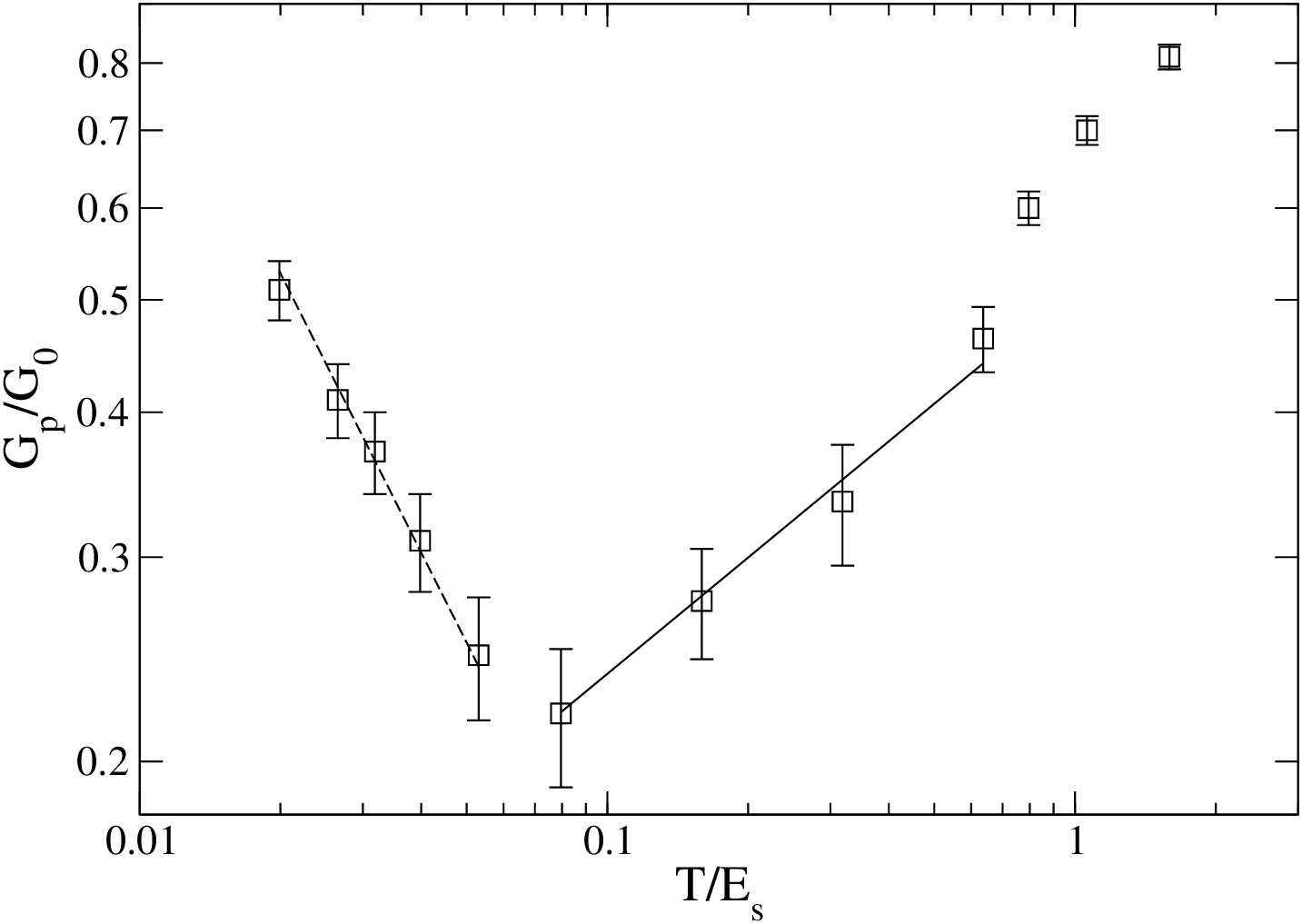}}
\caption{\label{fig4} 
Temperature dependence of the peak conductance $G_p$ for $g=0.6$,
$V_0/D=0.2$, and $E_s/D=\pi/20$. The dashed line reflects a 
$T^{-0.8}$ power law, the solid line a $T^{1/3}$ law.
}
\end{figure}

Next we discuss stronger barriers, $V_0/D=0.2$,
where one expects  sequential tunneling 
at not too low temperatures.  In Fig.~\ref{fig4},
data for the temperature dependence of the conductance peak height, $G_p(T)$,
is shown for $g=0.6$.
At low temperatures, $G_p$ approaches the perfect quantum conductance
 $G_0$ expected in the resonant tunneling regime. With increasing temperature,
the conductance goes through a minimum.
For intermediate temperatures, but still well below $E_s$, 
sequential tunneling then starts to dominate, and $G_p(T)$ {\sl increases} 
in a power-law fashion, $G_p(T)\propto T^\eta$.
Remarkably, our data are consistent with the CST scaling \cite{thorwart},
$\eta=2/g-3=1/3$, but not with the corresponding UST law 
\cite{furu98}, $\eta=1/g-2=-1/3$, which would even have a different sign.
We stress that other exponents, e.g. $\eta=2/g-2 =4/3$ as expected
for the effective {\sl strong} single-impurity problem arising at $T\gg E_s$
are incompatible with this numerically exact result.
Therefore our simulation data suggests that additional mechanisms beyond
conventional sequential tunneling can indeed be crucial.
As a side remark, we mention that
 Fig.~\ref{fig4} shows similar high-temperature ($T\approx E_s$)
features as the experimental data of Ref.~\cite{postma}.
The power-law upturn 
for $G_p(T)$ towards the resonant tunneling limit,
 see Fig.~\ref{fig4} at $T/E_s<0.1$,
has not yet been observed experimentally \cite{postma}, probably
due to a larger barrier height $V_0$ in the samples.

To conclude, we have developed and applied a 
numerically exact and well-controlled real-time Monte Carlo
approach to the computation of the resonant tunneling conductance
in a Luttinger liquid.  
For weak barriers and not too low temperature,
 we identify a Fabry-Perot regime, where
the conductance peak has a lineshape given by Eq.~(\ref{cond1})
with a temperature-dependent linewidth parameter $w_g(T)$. Within the range of
applicability, $w_g(T)$ exhibits $g$-dependent power-law scaling. 
At sufficiently low temperatures, 
for $g>1/4$, we then find a crossover into the universal coherent
resonant tunneling regime.   At higher temperatures, 
we find numerical evidence  in
support of the recently suggested 
correlated sequential tunneling picture \cite{thorwart}.
Finally, we note that
this algorithm also allows to study out-of-equilibrium
transport, quantum shot noise, and transport in the presence
of a Kondo-type dot.

\acknowledgments
We thank A.O. Gogolin, M. Grifoni, A. Komnik, V. Meden, K. Sch\"onhammer, 
and M. Thorwart for useful discussions.
This work has been supported by the EU and by the DFG.

\end{document}